\documentclass{article} 
\usepackage{iclr2022_conference,times}

\usepackage{amsmath,amsfonts,bm}

\def\eqref#1{equation~\ref{#1}}

\def\1{\bm{1}}

\DeclareMathAlphabet{\mathsfit}{\encodingdefault}{\sfdefault}{m}{sl}
\SetMathAlphabet{\mathsfit}{bold}{\encodingdefault}{\sfdefault}{bx}{n}

\usepackage{hyperref}
\usepackage{url}

\usepackage{colortbl, graphicx, multicol, soul, tipa}

\title{Reproducible Subjective Evaluation}

\author{Max Morrison\thanks{This material is based upon work supported by the National Science Foundation Graduate Research Fellowship under Grant No. DGE-1842165.}, Brian Tang, Gefei Tan \& Bryan Pardo\\
Northwestern University \\
\texttt{morrimax@u.northwestern.edu}\\
}

\iclrfinalcopy 
\begin{document}

\maketitle


\begin{abstract}


Human perceptual studies are the gold standard for the evaluation of many research tasks in machine learning, linguistics, and psychology. However, these studies require significant time and cost to perform. As a result, many researchers use objective measures that can correlate poorly with human evaluation. When subjective evaluations are performed, they are often not reported with sufficient detail to ensure reproducibility. We propose Reproducible Subjective Evaluation (ReSEval), an open-source framework for quickly deploying crowdsourced subjective evaluations directly from Python. ReSEval lets researchers launch A/B, ABX, Mean Opinion Score (MOS) and MUltiple Stimuli with Hidden Reference and Anchor (MUSHRA) tests on audio, image, text, or video data from a command-line interface or using one line of Python, making it as easy to run as objective evaluation. With ReSEval, researchers can reproduce each other's subjective evaluations by sharing a configuration file and the audio, image, text, or video files.
\end{abstract}


\begin{figure}[h]
		\centering
		\includegraphics[width=\textwidth]{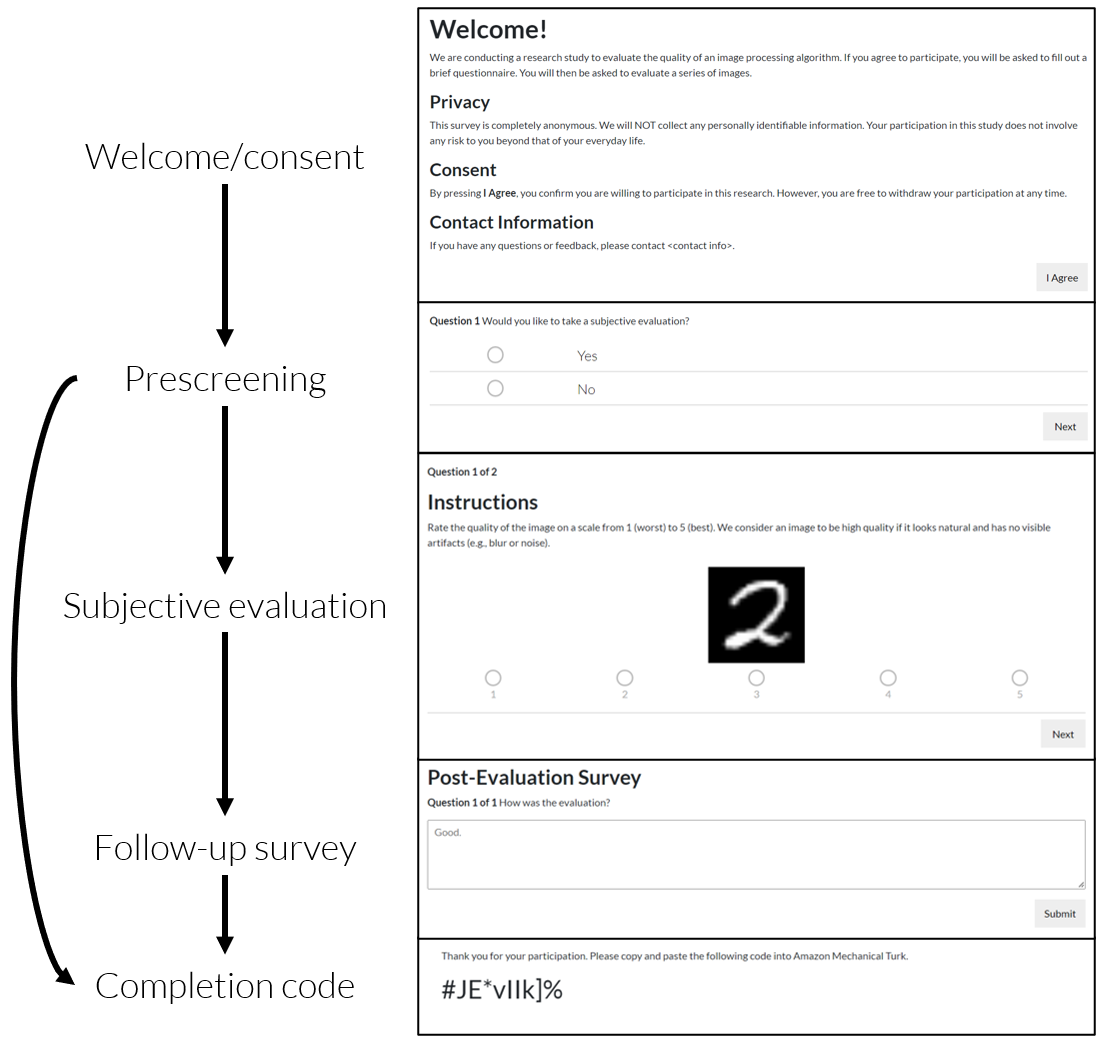}
		\caption{An overview of the stages of a subjective evaluation on ReSEval. After obtaining the consent of the participant, we perform a prescreening step. Participants who pass the prescreening step perform a subjective evaluation and take a follow-up survey. Participants who complete the follow-up survey or do not pass the prescreening are taken to the final page, which provides a completion code for participants to enter on the crowdsource platform to receive payment for their work. All of the text is configurable via Markdown. The type of test, type of question (free response or multiple choice), and number of questions are also configurable.}
		\label{fig:system}
\end{figure}

\section{Introduction}
\label{sec:intro}

Subjective human evaluations of audio, image, text, or video data are a standard evaluation methodology in machine learning~\citep{sheng2019machine}, linguistics~\citep{cole2017crowd}, and psychology~\citep{kazdin2021research}. For example, a machine learning researcher may want to know whether a conditionally generated image corresponds to its conditioning ~\citep{ramesh2021zero, shoshan2021gan}, and a linguist may be interested in measuring how humans perceive the pitch of speech~\citep{cole2017crowd}.

Recruiting a large number of participants for in-person evaluations is a time-consuming process. For some types of tasks, comparable results can be achieved by performing evaluation on a crowdsourcing platform, such as  Amazon Mechanical Turk (MTurk) or Prolific. As long as the evaluation is designed properly~\citep{cartwright2016fast, sai2020survey}, such evaluations can be effective substitutes for in-person evaluations. However, researchers are often faced with strict page limits for conference submissions, and details of the evaluation important for replicating the study are routinely omitted from published papers. Frequently omitted  details include the specifics of the methods used to prescreen participants, the wording of the instructions, and the type of sampling used to assign evaluation files to participants. 

Even in the case where something like a prescreeing test is described in the publication, it can still be difficult to reproduce. A prescreening test is often performed to establish whether participants have the perceptual acuity to perform an evaluation task (e.g., passing a hearing test prior to evaluating a text-to-speech system). While crowdsourcing platforms such as MTurk offer some ability to filter participants based on qualifications, perceptual screening tests ~\citep{cartwright2016fast, woods2017headphone} are typically not available in the default platforms, requiring researchers to develop and deploy their own prescreening code, which others would need access to if they are to truly reproduce the work.

While MTurk contains many easy-to-use evaluation templates for some tasks, as of March 2022 they do not support many common research tasks in machine learning (e.g., A/B tests of images for image super-resolution, ABX tests of text for continuation tasks, or MUSHRA tests of audio quality). Finally, for machine learning researchers, platforms such as MTurk do not integrate easily into the typical Python-based software development workflow. 

Given the issues associated with performing high-quality, reproducible subjective evaluation, many researchers have proposed proxy metrics that can be used to estimate human perception. However, whenever such an objective measure is proposed, subsequent research demonstrates its failures. For example, Frechét Inception Distance (FID)~\citep{NIPS2017_8a1d6947}, a common objective metric for image generation, is biased towards its training data and is not aligned with human perceptions~\citep{jung2021internalized}. Peak Signal-to-Noise Ratio (PSNR) is frequently used for image and vision denoising~\citep{zhang2020comparing}, but weights each pixel as equally important to visual perception and is not able to measure the perceptual impact of various types of distortions~\citep{wang2009mean}. Metrics for natural language understanding such as BLEU~\citep{papineni2002bleu} have been unable to produce even a moderate correlation with human evaluations of nautral language generation systems~\citep{novikova2017we, sai2020survey}. In speech synthesis, the DeepSpeech Distances~\citep{binkowski2019high} of two state-of-the-art systems do not correlate with human judgments obtained via A/B testing~\citep{morrison2021chunked}. Some works propose to learn an objective metric from human subjective evaluation data. For example, the Learned Perceptual Image Patch Similarity (LPIPS) is trained on A/B tests of image patches~\citep{zhang2018unreasonable}, and the Contrastive Deep Perceptual Audio Metric (CDPAM) is trained on an A/B test of speech quality~\citep{manocha2021cdpam}, but the correlation of such measures with human perception is only guaranteed on the range of data on which the systems are trained.

What is needed is an approach to subjective evaluation that is truly reproducible and integrates well with machine learning development workflows. To this end, we propose Reproducible Subjective Evaluation (ReSEval), an open-source framework for performing crowdsourced subjective evaluations. ReSEval allows researchers to launch a subjective evaluation either from the command line or using one line of Python, making it as easy to run as objective evaluation. This enables complex machine learning evaluation pipelines, such as automatically evaluating whether a new generative model should replace a production model, or automatically determining when to stop training a model based on human preference scores. Further, ReSEval offers researchers a convenient way to make their subjective evaluations reproducible by their research community by simply releasing the configuration file and evaluation files they used when performing evaluation with ReSEval.

Papers-with-code and open GitHub repositories have provided a much-needed boost to machine learning research reproducibility. Deployment of trained models in online repositories, such as HuggingFace, has been another important step. ReSEval is an essential component to take the next step forward in reproducible research. Imagine online challenges or Papers-with-Code leader boards where all submissions are automatically evaluated using the exact same subjective study evaluation. Imagine being able to verify all the details of a study design because its exact implementation was posted on GitHub. ReSEval makes this possible. ReSEval is available under an open-source license at
\texttt{\href{https://github.com/reseval/reseval}{github.com/reseval/reseval}}.


\section{Best practices for crowdsourced subjective evaluations}

We next discuss best practices for performing crowdsourced subjective evaluation, focusing on best practices that are included by default or simple to implement with ReSEval. For a more thorough treatment of best practices, see ~\citep{mason2012conducting}.

\textbf{Use the correct test} By default, ReSEval includes many of the most widely-used subjective evaluation test paradigms, including A/B, ABX, MOS, and MUSHRA-style tests.

\begin{itemize}
    \item \textbf{A/B} - The participant is asked which of two audio, image, text, or video examples ranks higher along a perceptual attribute (e.g., audio quality or text sentiment).
    \item \textbf{ABX} - The participant is asked which of two stimuli (e.g. two images) is, e.g., more similar to a reference stimulus (a canonical image).
    \item \textbf{Mean opinion score (MOS)} - The participant rates a perceptual quality of an audio, image, text, or video example from 1 (worst) to 5 (best).
    \item \textbf{MUltiple Stimuli with Hidden Reference and Anchor (MUSHRA)} - The participant is presented multiple audio, image, text, or video examples and uses sliders to rate each of the examples.We refer to these as MUSHRA-style, rather than MUSHRA tests, because MUSHRA denotes a specific standard audio evaluation methodology, which is detailed in the International Telecommunications Document BS.1534~\citep{liebetrau2014revision}
\end{itemize}

It is important that the hypothesis being tested is consistent with the test format. For example, while MOS and MUSHRA-style tests are conveniently capable of evaluating more than one pair of conditions at once, they are less able to demonstrate a statistically significant difference between two conditions when the perceptual differences between those conditions is subtle relative to other conditions. An A/B test can expose these subtleties as statistically significant, even with many fewer participants. 

This has implications for the design of ablation studies in machine learning: if the ablation conditions are significantly worse than the proposed model, an MOS or MUSHRA-style test that includes the ablation conditions, the proposed model, and a ground-truth reference will skew ratings of the proposed model towards ground-truth. For an example of this in the field of speech synthesis, see~\citep{kong2020hifi} and~\citep{morrison2021chunked}. However, MOS tests can be a cost-effective way of comparing many conditions at once. Therefore, one cost-effective way to show statistical significance is to first conduct an MOS (or MUSHRA-style) test, and then perform A/B tests on pairs of conditions where the test shows an inconclusive preference. Usually, this process requires researchers to manually review results and deploy a second user study. ReSEval makes it possible for the researcher to specify in code when to automate the deployment of, e.g., an A/B test, conditional on the results of a prior test (e.g., an MOS test).

MOS tests are typically cheaper than MUSHRA-style tests as they require less time for participants to complete. However, they also have limited ability to rank-order samples relative to MUSHRA-style tests, which can be useful data for downstream tasks, such as training a machine learning model to approximate human subjective preference. For both MOS and MUSHRA-style tests, using high- and low-anchor conditions can minimize changes in the range and variance of the results between evaluations with different conditions. These anchors are representative conditions of very good and bad examples for the specific task being evaluated. For example, text-to-speech uses ground-truth recorded speech as the high-anchor, and might use speech with added noise as a low-anchor.

\begin{figure}[t]
		\centering
		\includegraphics[width=\textwidth]{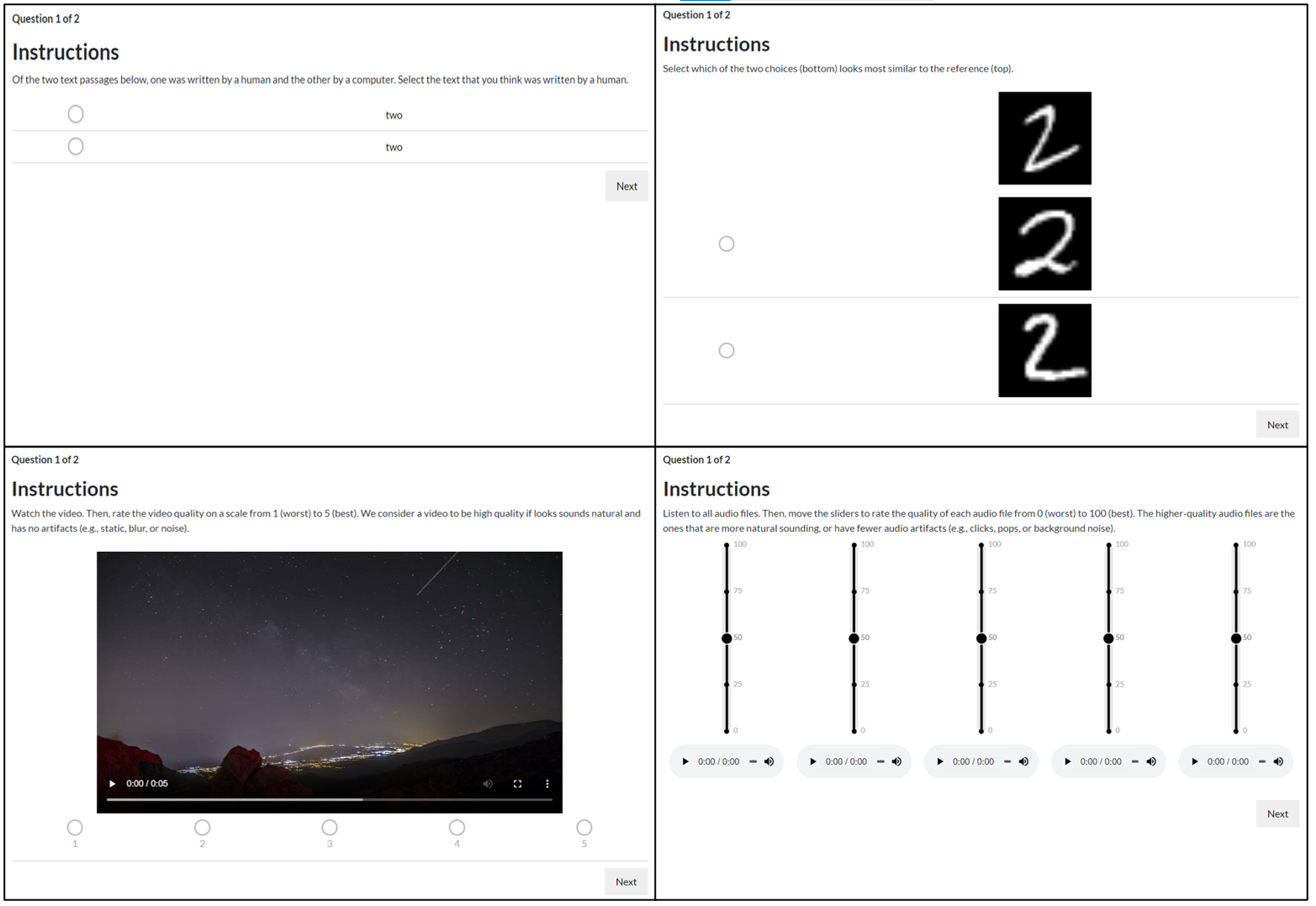}
		\caption{Four examples of subjective evaluation tests available with ReSEval. \textbf{(Top left)} An A/B test with text data. \textbf{(Top right)} An ABX test with images. \textbf{(Bottom left)} An MOS test on videos. \textbf{(Bottom right)} An audio MUSHRA test.}
		\label{fig:tests}
\end{figure}

\textbf{Prescreen} It is important that participants be well-qualified to perform evaluation. Depending on the specific evaluation, this can include fluency in a particular language, having no vision impairment, or having access to headphones and a quiet listening environment. Crowdsource platforms permit filtering participants by some of these criteria. Other filtering criteria require prescreening to be included as part of the survey design. Reproducing this prescreening step is critical for reproducing a subjective evaluation. Deploying a study through ReSEval makes perfect reproduction of the prescreening simple, as the full specification is included.

\textbf{Account for learning curves and fatigue effects} The order in which samples are presented to new participants as well as the number of samples presented can impact ratings~\citep{schwarz2016effects}. For example, while high-anchors and low-anchors can minimize changes in the range and variance of the results for MOS tests, a new participant performing their first evaluation question has no basis for comparison. Showing participants high- and low-anchor examples prior to evaluation can reduce this learning curve by pre-establishing a perceptual range. As well, it is important to limit the total length of the evaluation and not allow participants to repeat the evaluation in order to prevent fatigue effects from affecting evaluation results.

ReSEval randomly orders examples for participants to prevent learning and fatigue effects from biasing specific examples. Given a random seed, these assignments are deterministic, meaning that any learning and fatigue effects are replicable if the study is repeated with the same seed.



\section{ReSEval}

We provide a brief overview of ReSEval\footnote{This paper describes ReSEval version \texttt{0.0.1} at commit \texttt{3843cdb}}. ReSEval is a Python package that permits A/B, ABX, MOS, and MUSHRA-style tests on audio, image, text, and video data. Crowdsourced participants in a subjective evaluation are first presented with an introduction screen that describes the test, followed by an optional prescreening step, the evaluation, and an optional followup survey (Figure~\ref{fig:system}). Users of ReSEval can configure all of the text of the introduction, prescreening, followup survey, and evaluation instructions in a single configuration file via Markdown.
This configuration also includes the type of test administered (e.g., A/B or MOS); the participant filtering criteria; the participant pay; and the cloud services used for file storage, database management, and server hosting. We designed this configuration to include all of the parameters necessary to fully replicate a crowdsourced subjective evaluation, provided access to the same audio, image, text, or video files being evaluated and the same version of ReSEval. After setup, a subjective evaluation can be launched in one line on the command-line given a configuration file \texttt{<config>} and a directory of evaluation files \texttt{<directory>}.

\begin{verbatim}
    python -m reseval <config> <directory> --production
\end{verbatim}

This command sets up storage, database, and server resources for a subjective evaluation, launches the subjective evaluation to crowdsourced participants, monitors progress, performs statistical analysis of results, pays participants, and shuts down compute resources once evaluation has finished. These steps can be performed individually for more control, and can also be called directly via our Python API. ReSEval can be run locally or in remote development mode (e.g., using the MTurk Sandbox) in order to debug evaluations before deployment. ReSEval runs on Linux, MacOS, and Windows and can be installed via pip\footnote{\texttt{pip install reseval}}.

\section{Conclusion} We present Reproducible Subjective Evaluation (ReSEval), a framework for performing reproducible crowdsourced subjective human evaluations as easily as objective evaluation. ReSEval lowers the barrier-to-entry for performing high-quality crowdsourced subjective evaluations, a necessary step for many research tasks in machine learning, linguistics, and psychology. We are continuing to add features to ReSEval, such as additional screening tests, a pre-test page that shows participants good and bad examples to reduce the initial learning curve, integrated support for performing attention checks, and support for additional cloud compute platforms (e.g., Amazon Web Services and Firebase) and crowdsourcing platforms (e.g., Prolific).


\bibliography{iclr2022_conference}
\bibliographystyle{iclr2022_conference}

\end{document}